# Impact of High PV Penetration on the Inter-area Oscillations in the U.S. Eastern Interconnection

Shutang You

*Abstract*—**This study explores the impact of high PV penetration on the inter-area oscillation modes of large-scale power grids. A series of dynamic models with various PV penetration levels are developed based on a detailed model representing the U.S. Eastern Interconnection (EI). Transient simulations are performed to investigate the change of inter-area oscillation modes with PV penetration. The impact of PV control strategies and parameter settings on inter-area oscillations is studied. This study finds that as PV increases, the damping of the dominant oscillation mode decreases monotonically. It is also observed that the mode shape varies with the PV control strategy and new oscillation modes may emerge under inappropriate parameter settings in PV plant controls.**

*Index Terms*— **Large-scale power system oscillation, photovoltaic (PV), oscillation mode, oscillation frequency, inertia**

## I. Introduction

Solar photovoltaic (PV) generation grows quickly in the U.S. and worldwide, joining wind power as a major renewable energy technology. While many aspects of wind power's impact on system power grid stability has been investigated, including rotor angle stability [1-3], voltage stability [4], frequency response [5-7], and inter-area oscillations [1, 8-11], the impact of PV generation on power system dynamics has not been given enough attention.

Among all these aspects, inter-area oscillations need special attention from system operators. Poorly damped inter-area oscillations can reduce transmission line capacity, damage system generation and transmission facilities, influence power quality, and lead to cascading failures or blackouts. The impact of PV generation on power system oscillations may be attributed to various aspects. However, the findings of existing literature are not consistent. For example, the New-England and New York test system was used to investigate the PV's impact on its established inter-area oscillation mode; it was found that PV could detrimentally affect the inter-area oscillation mode as PV integration results in larger angular separation among synchronous generators [12]. Furthermore, using small signal analysis and transient simulation, the study in [13] indicated that the increase of utility-scale and residential rooftop PV may decrease damping of inter-area oscillation modes due to reduced system inertia. However, other studies have found it is also possible that PV integration could improve small-signal stability. For example, the study in [14] concluded that PV could increase oscillation damping because PV adds damping to critical modes, and the scattered integration pattern is more beneficial than the concentrated pattern. Other studies that have similar findings include [15] and [16]. Additionally, other literature observed both beneficial and detrimental influence of PV on oscillations. Ref. [17] studied the impact of increased PV on system small signal stability based on a single-machine infinite-bus system. It was found that PV could either have positive or negative impact on oscillation damping. Similar conclusions can be found in [11] and [18]. The operation limit of PV also plays a significant role on the contribution of PV's damping torque. Ref. [19] studied the impact of distributed PV and large PV farms on system stability and found that damping does not vary significantly with the increase of either PV type.

Obviously, the impact of PV on power system inter-area oscillations has not been well understood yet. Since there are so many factors that may play a role in this mechanism, the impact of PV on inter-area oscillations may need to be studied on a case-by-case basis at this stage. Therefore, the United States Eastern Interconnection (EI) was used as a case study in this paper to investigate the impact of PV generation on a large interconnected power grid. Specifically, by incrementally displacing synchronous generators with PV in the EI dynamic model, how the dynamics of PV affect inter-area oscillation frequency and damping ratio will be illustrated. The following sections present the procedures and details in model development, time-domain dynamic simulation, and analytical analysis.

## II. Oscillation Mode Analysis in Current EI System

Since the EI system covers a wide area, inter-area oscillation is important issue for system operators and planners [20, 21]. Our previous study shows that the inter-area oscillation modes are observable in FNET/GridEye, which is a wide-area measurement system deployed at the distribution level. Before analyzing the impact of PV on oscillation of the U.S. EI system, this section analyzes the characteristics of the inter-area oscillation in the EI system based on the FNET/GridEye wide-area measurements during 2013 to 2015 using the Matrix-Pencil method [22]. Fig. 1 shows the frequency distribution of the dominant oscillation mode, while Fig. 2 shows the distribution of the damping ratio of the dominant oscillation mode. It can be seen that the frequency of the dominant oscillation of EI is around 0.2 Hz and the center of damping ratio is close to 10%. Additionally, both two distributions show a distribution pattern with long tails, similar to the Beta distribution. The slight shifting of oscillation frequency and damping can be explained by the seasonal and daily load variations.

---

Table I shows the oscillation frequency and damping ratio of different years. It can be seen that the oscillation characteristics are relatively stable in the past years. The frequency center changed slightly from 2014 to 2015 (0.21Hz to 0.20Hz). It can be explained by the increase of renewable generation in EI system. As the 0.2 Hz mode is the dominant oscillation in the EI system, this mode will be the study focus in the rest of this paper.

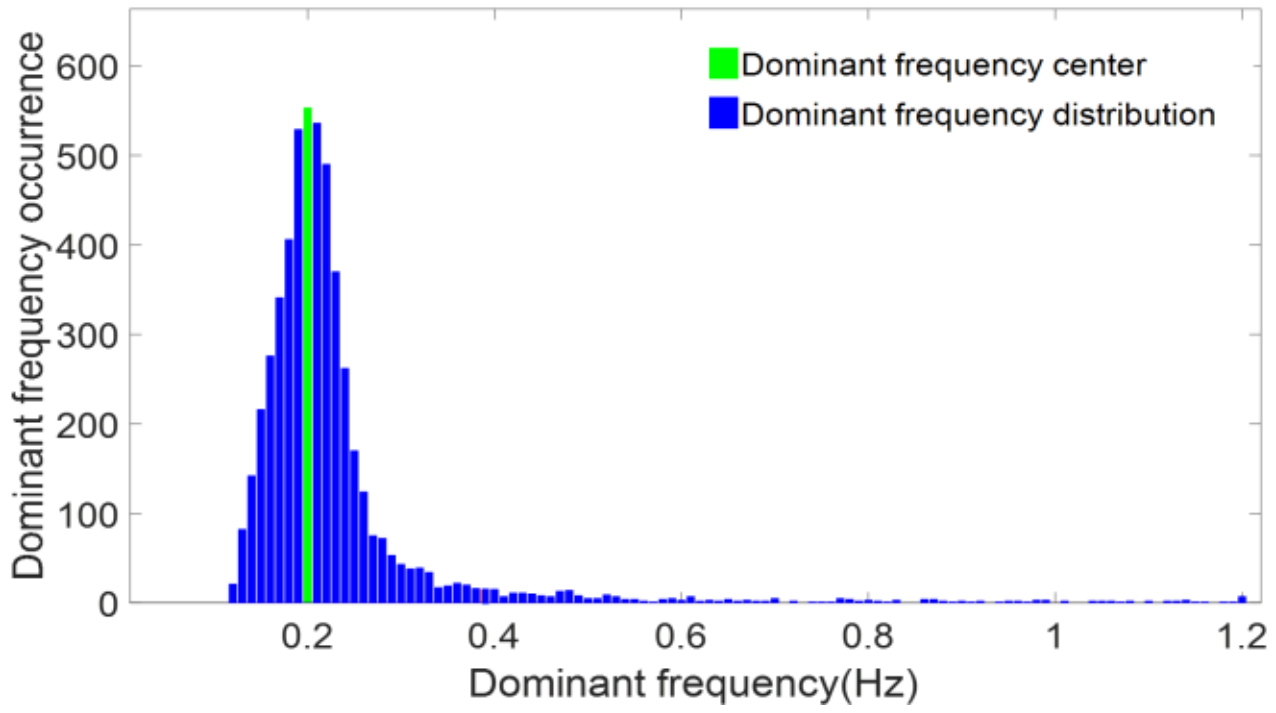

Fig. 1. Dominant frequency distribution of inter-area oscillations.

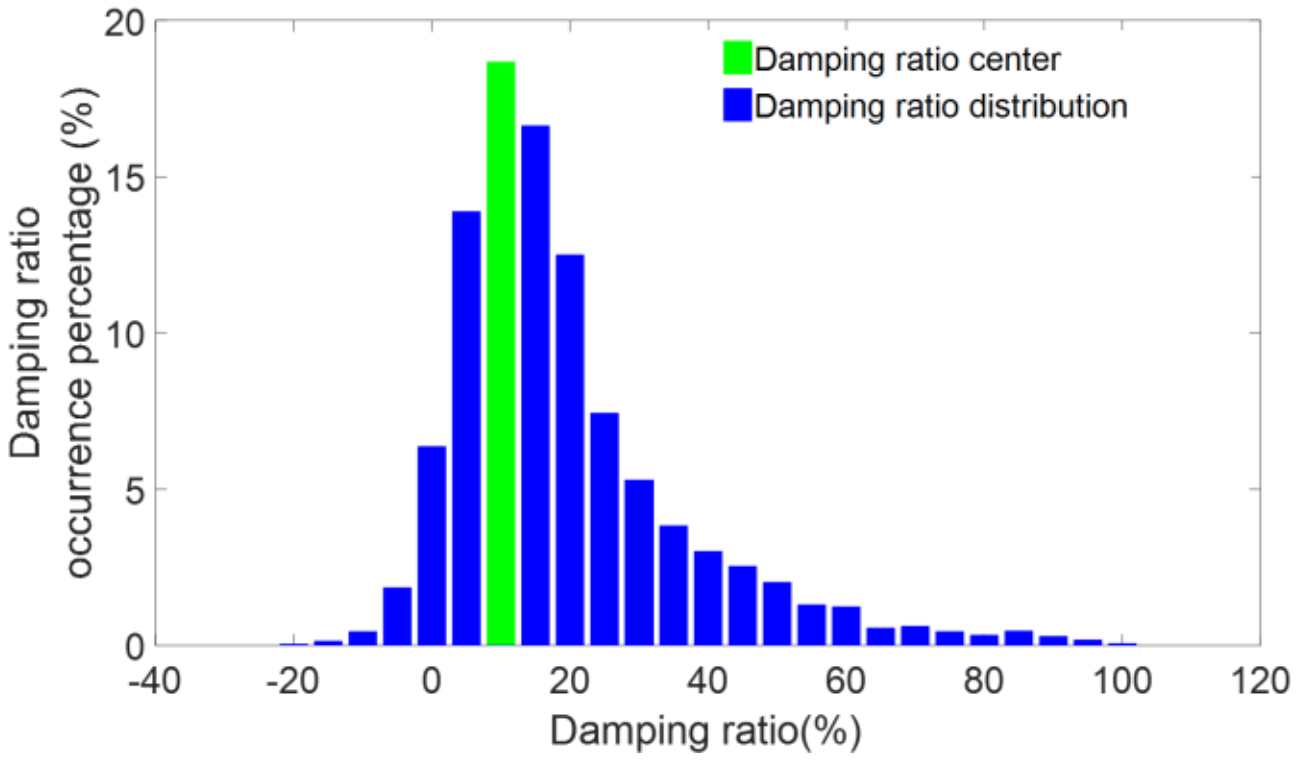

Fig. 2. Damping ratio distribution of inter-area oscillations

TABLE I. OSCILLATION FREQUENCY AND DAMPING RATIO

| Year | 2013 | 2014 | 2015 |
|---|---|---|---|
| Dominant frequency center (Hz) | 0.21 | 0.21 | 0.20 |
| Damping ratio center of the dominant frequency (%) | 10 | 10 | 10 |

## III. HIGH PV EI SYSTEM MODELLING APPROACH

High PV penetration. model development, time-domain transient simulation, and oscillation analysis will be conducted to investigate the change of the 0.2Hz oscillation mode with the increase of PV. In this section, the high PV penetration model development will be introduced.

### *A. Base Model Introduction*

This study is based on the detailed power flow scenarios for the EI system in year 2030, which was originally developed by the Eastern Interconnection Planning Collaborative (EIPC) [23]. This power flow model depicts a scenario in which wind generation provides 15% instantaneous power and some transmission network upgrades are carried out to facilitate wind power transferring primarily from the west EI to the east EI. Earlier efforts also include the development of corresponding dynamic models, which includes synchronous generators, excitation systems, turbine governors, and load dynamic models. Extensive sanity check and contingency simulation are conducted to ensure numerical convergence and simulation accuracy. More detailed information on dynamic modeling can be found in [24]. Some statistic values on this model are shown in Table 1,

### *B. High PV Penetration Model Development*

Before developing the high PV power system models, this study defines the PV penetration rates for the to-be-developed high PV simulation scenarios. Based on the result of a survey involving electric utilities, national labs and research institutes, the renewable generation mix of the to-be-developed high PV simulation scenarios are defined to be 5%, 25%, 45%, and 65% for PV penetration (as shown in Table II), plus 15% wind at the interconnection level [25]. The rest is conventional generation including hydro, nuclear, and thermal power plants.

The development of high PV penetration model includes two major steps: 1) Generate PV distribution. 2) Incorporate the dynamic models of PV and wind into the system model.

In the first step, the guideline of PV plant siting is to optimize the distribution of PV based on various impact factors, including existing generation and transmission infrastructure, load forecasting, solar radiation, fuel price forecast, carbon emission, PV price forecast, and the PV siting land price, etc. The horizon of the PV projection features a PV growth primarily driven by a high carbon-emission price curve as predicted in [26]. A summary of the sources of input data in the PLEXOS optimization model[1] is shown in Table III [27]. Using these parameters as inputs, a PV projection model is applied to optimize the PV distribution to minimize the total cost in the projected high PV scenarios.

TABLE I. BASIC INFORMATION OF THE EI MODEL

| **EI model statistics** | **Value** |
|---|---|
| Total bus number | 68309 |
| Generator number | 8337 |
| Branch number | 58784 |
| Load | 560 GW |

TABLE II. PV AND WIND PENETRATION RATES OF ALL SCENARIOS IN EI.

| Scenario | Instantaneous PV penetration level |
|---|---|
| Scenario 1 | 5% |
| Scenario 2 | 25% |
| Scenario 3 | 45% |
| Scenario 4 | 65% |

[1] It uses a bubble/pipe model representing 24 partitioned regions in EI and the interfaces between them These regions represent utilities, regional transmission operators, coordinating authorities, independent system operators or other natural groupings based on the structure of the EI.

TABLE III. PLEXOS MODEL INPUT DATA SOURCES

| PLEXOS model input | Data sources |
|---|---|
| • Existing generation and transmission infrastructure<br>• Load forecast<br>• Solar radiation<br>• Fuel price forecast<br>• Carbon emission price forecast | The Eastern Interconnection Planning Collaborative (EIPC) dataset[2] [26, 28-30] |
| • PV price forecast | North American PV Outlook [31] |
| • PV siting land price | Land Value 2015 Summary [32] |

The optimization problem is solved by a commercial mix-integer programming solver (Xpress-XP). The optimization result is the projected PV capacity distribution in each region for each PV penetration levels.

The PV distribution in the four renewable penetration levels are shown in Fig. 4. Following the PV distribution optimization, the second step is to incorporate PV dynamic models into the system for each scenario. The connectivity diagram of the PV dynamic model is shown in Fig. 3 [33]. In this step, the distribution of PV is kept consistent with the PV siting optimization result. As the high PV power systems are future scenarios, generic dynamic parameters of PV power plants are adopted in the high PV model [33]. Table IV shows the two typical control strategies of PV power plants [34]. According to the current PV plants settings in North America., the Volt/Var control with SolarControl (Strategy 1) is selected as the base strategy.

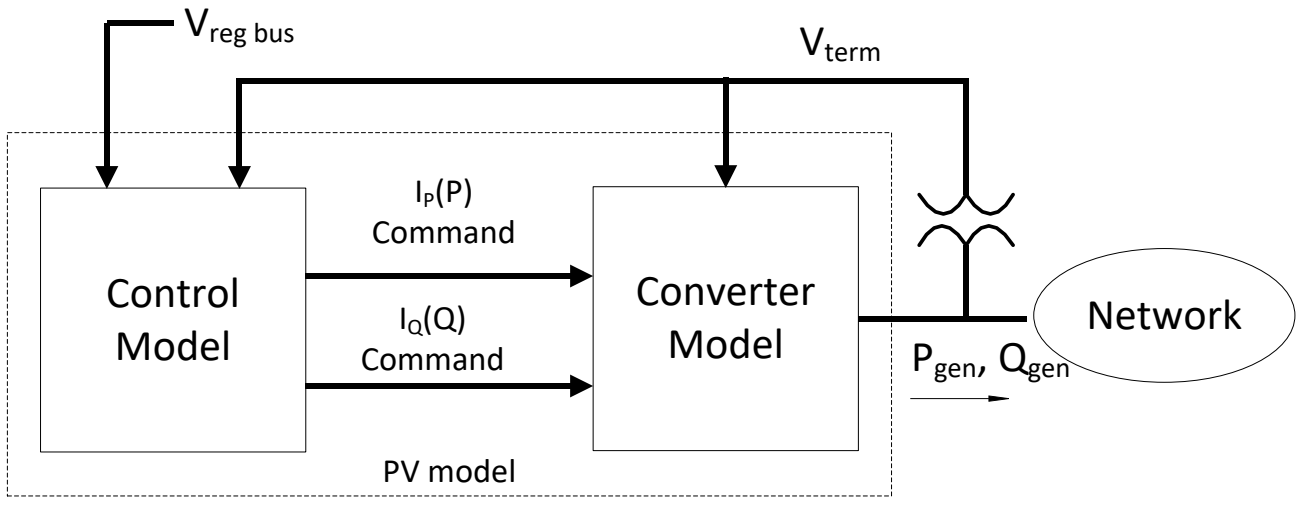

Fig. 3. PV dynamic model connectivity

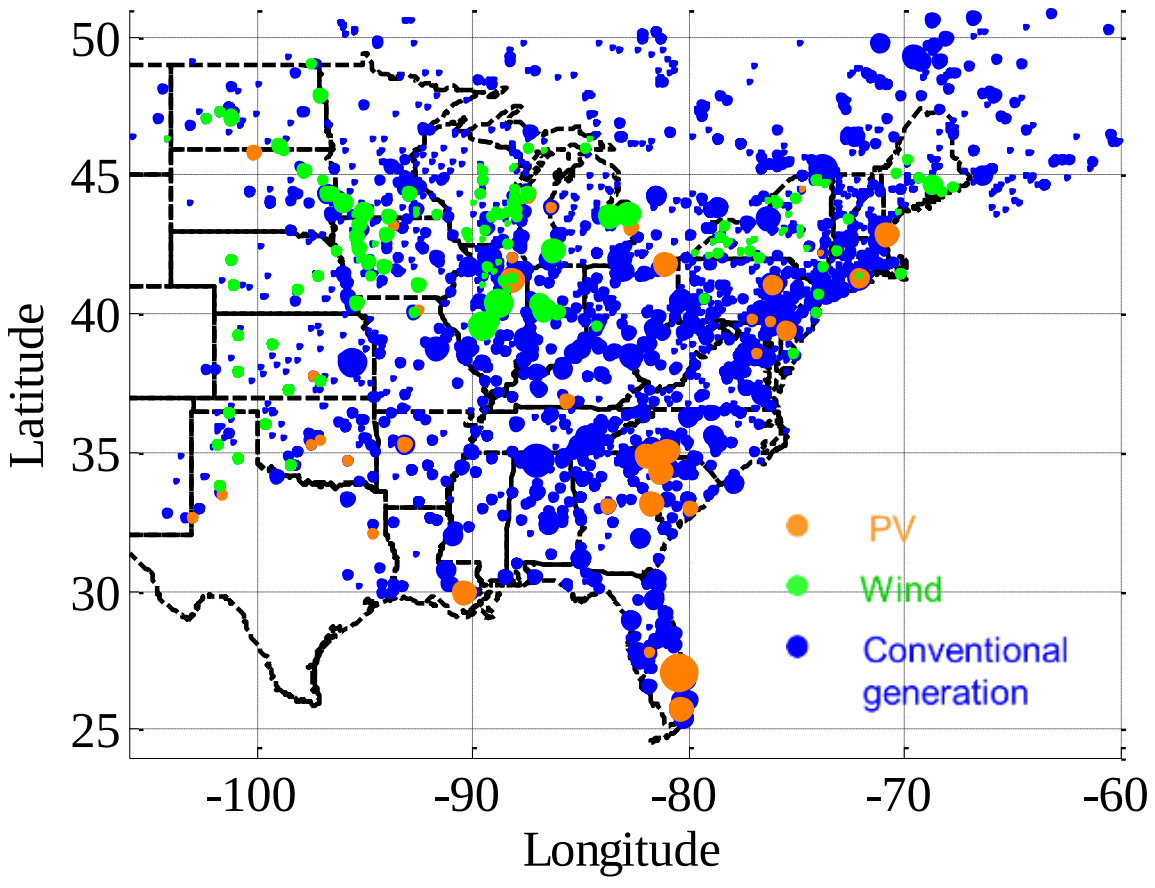

(a) 5%PV + 15%WT

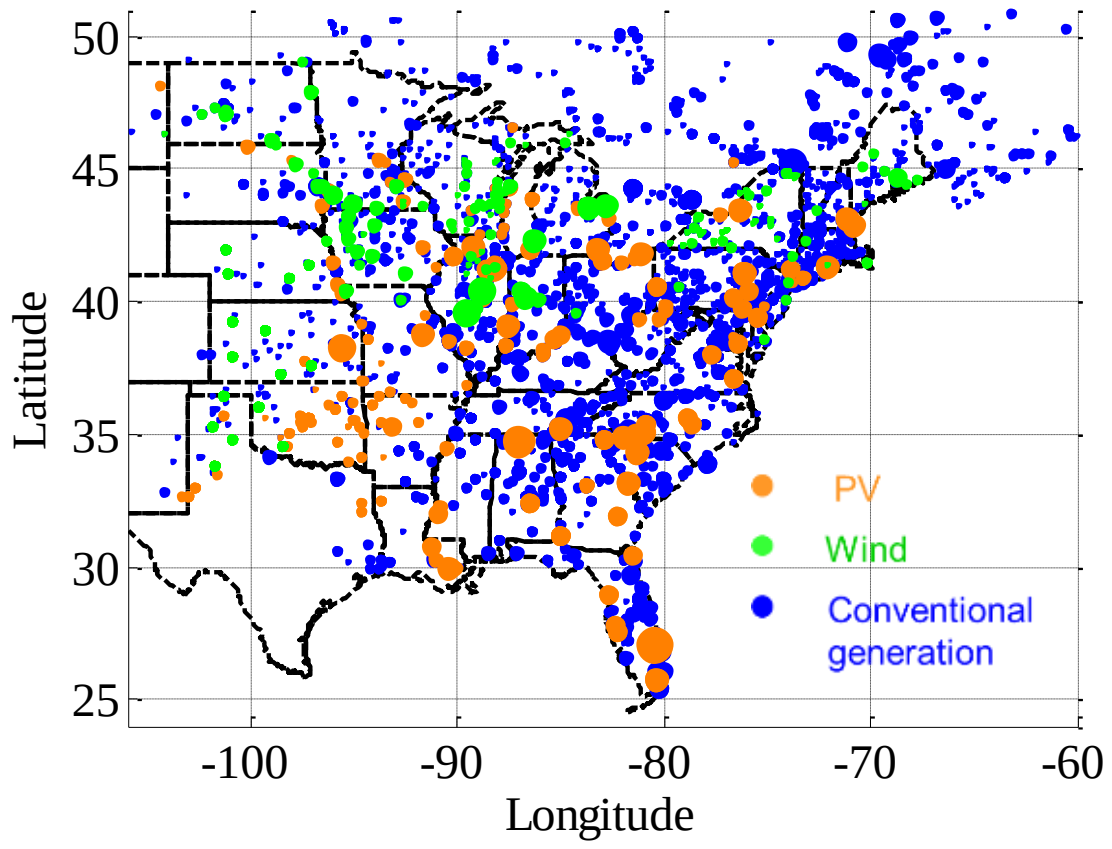

(b) 25% PV + 15%WT

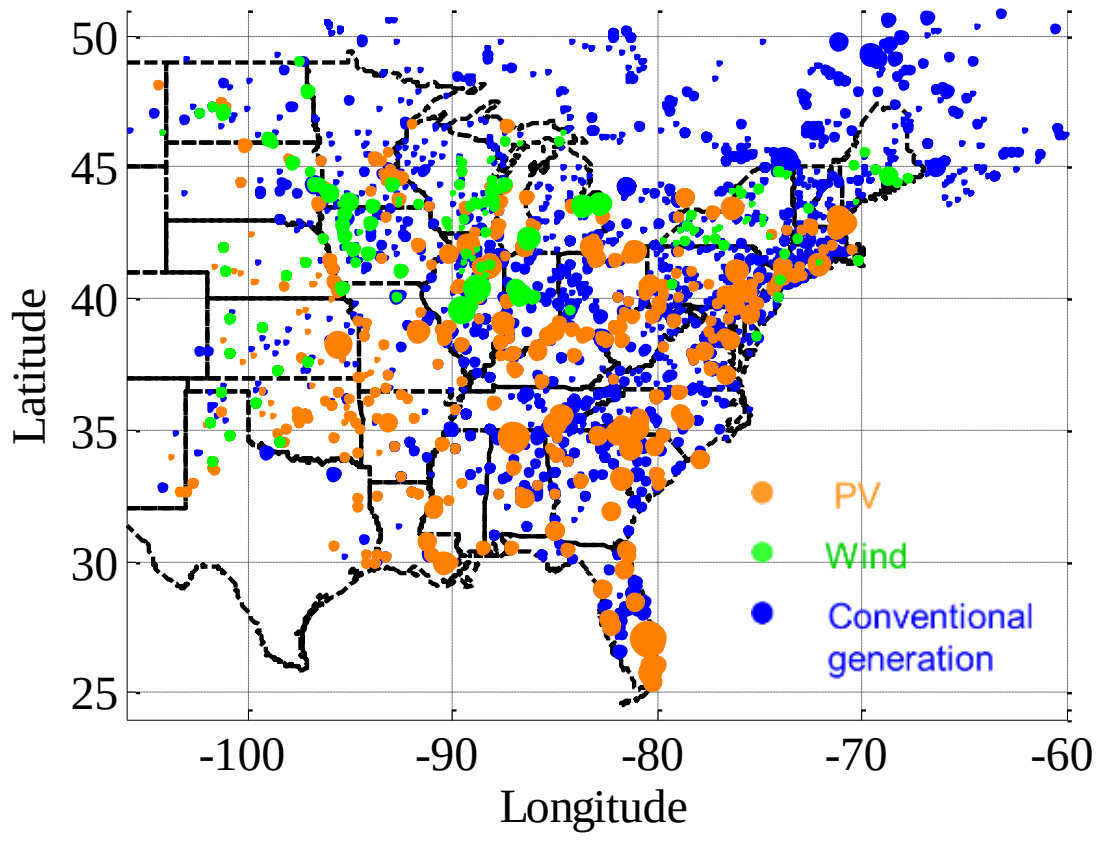

(c) 45%PV+15%WT

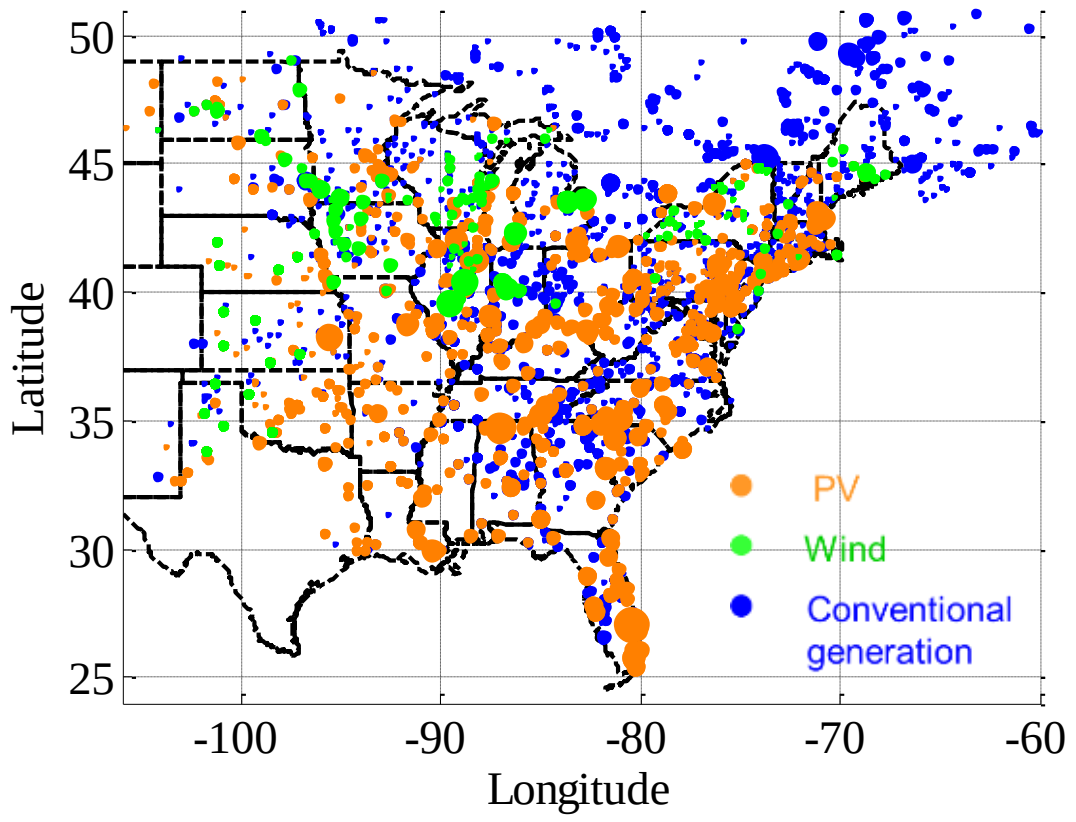

(d) 65%PV + 15%WT

Fig. 4. PV distribution in different scenarios

TABLE IV. PV PLANTS CONTROL STRATEGIES [33]

| PV control strategies | Description |
|---|---|
| Strategy 1: Volt/Var control with SolarControl[3] | Current configuration of North American PV plants |
| Strategy 2: Volt/Var control without SolarControl | SolarControl turned off and there is a slow reset of reactive power; terminal voltage control is rapid |

## IV. IMPACT OF HIGH PV ON EI SYSTEM OSCILLATION

The EI system is a geographically dispersed power grid. To capture its inter-area oscillation modes, multiple observation locations across the EI are necessary to mitigate the impact of location oscillations and obtain more reliable results of inter-area oscillation analysis. The observation locations in this study are shown in Fig. 5. To analyze the oscillation observed at each observation point, the Matrix Pencil method is applied to the bus frequency at each observation location [35]. The oscillation analysis results at all observation locations are combined to form a more complete picture of oscillation. These procedures are conducted under four scenarios, namely 5%, 25%, 45%, and 65% PV penetration.

[2] The EIPC dataset is the main dataset in optimization. This dataset was created by Charles River Associates (CRA) in an effort supported by U.S. Department of Energy involving major stakeholders and planning coordinators in the Eastern Interconnection. Energy Exemplar translated this dataset into PLEXOS for multiple EI expansion studies

[3] The SolarControl function monitors a specified bus voltage and compares it to a reference voltage.

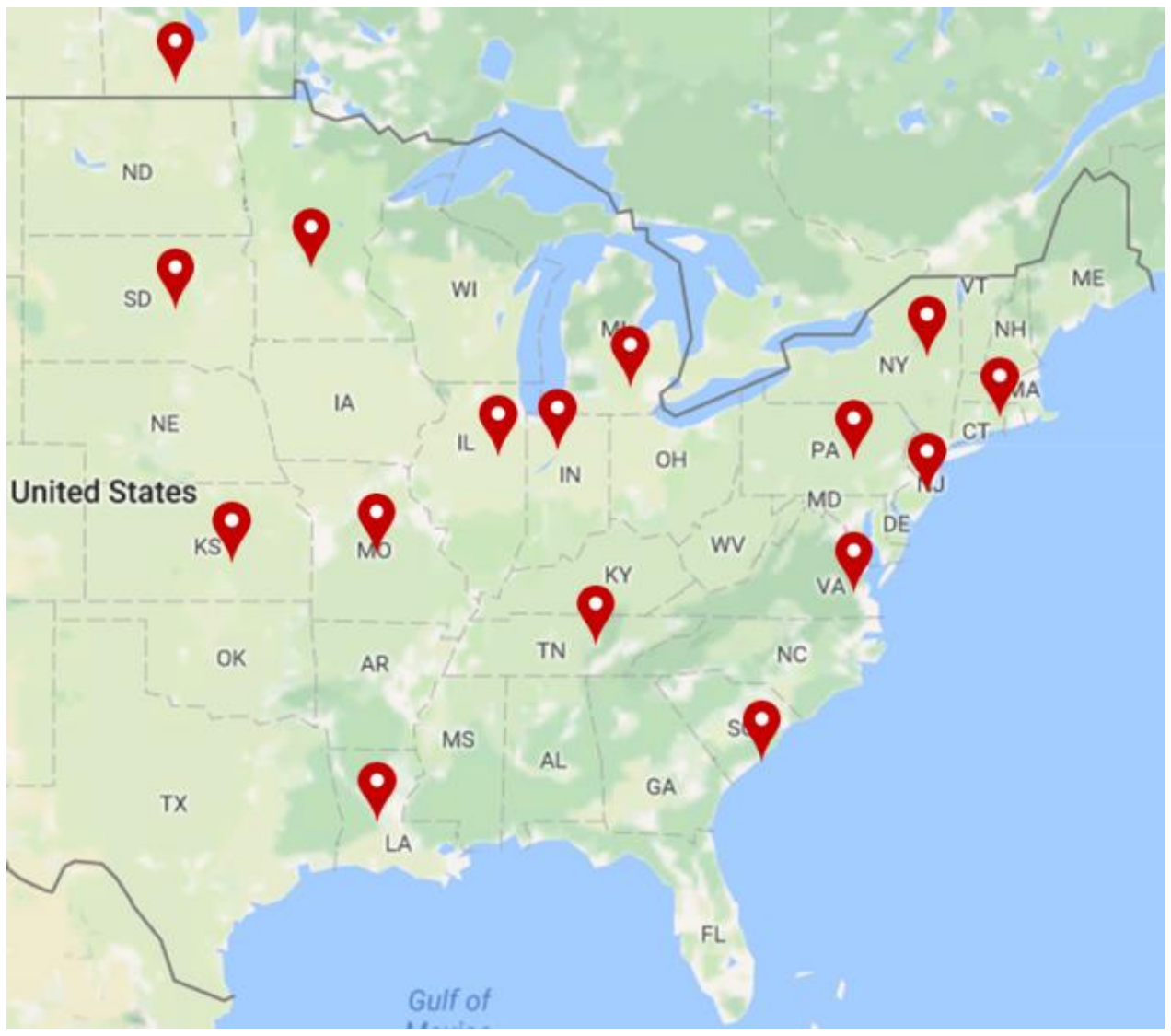

Fig. 5. Observation locations in the U.S. EI system

## A. *Impact on Frequency and Damping Ratio*

To observe the oscillation, a test disturbance is applied to the system and the frequency responses are recorded at multiple locations. This disturbance is a three phase fault on a 500kV bus located in the central EI, lasting for two cycles. Using Matrix Pencil analysis to analyze the oscillation mode at multiple locations, the frequency and damping ratio of the 0.2Hz inter-area oscillation mode at each observation location are calculated. The change of oscillation frequency and damping ratio with PV penetration are shown in Fig. 6 and Fig. 7, respectively. It can be noted that as PV penetration increases from 5% to 65%, the oscillation frequency increases almost linearly from 0.20Hz to 0.28Hz, and the damping ratio decreases from around 9% to 6%.

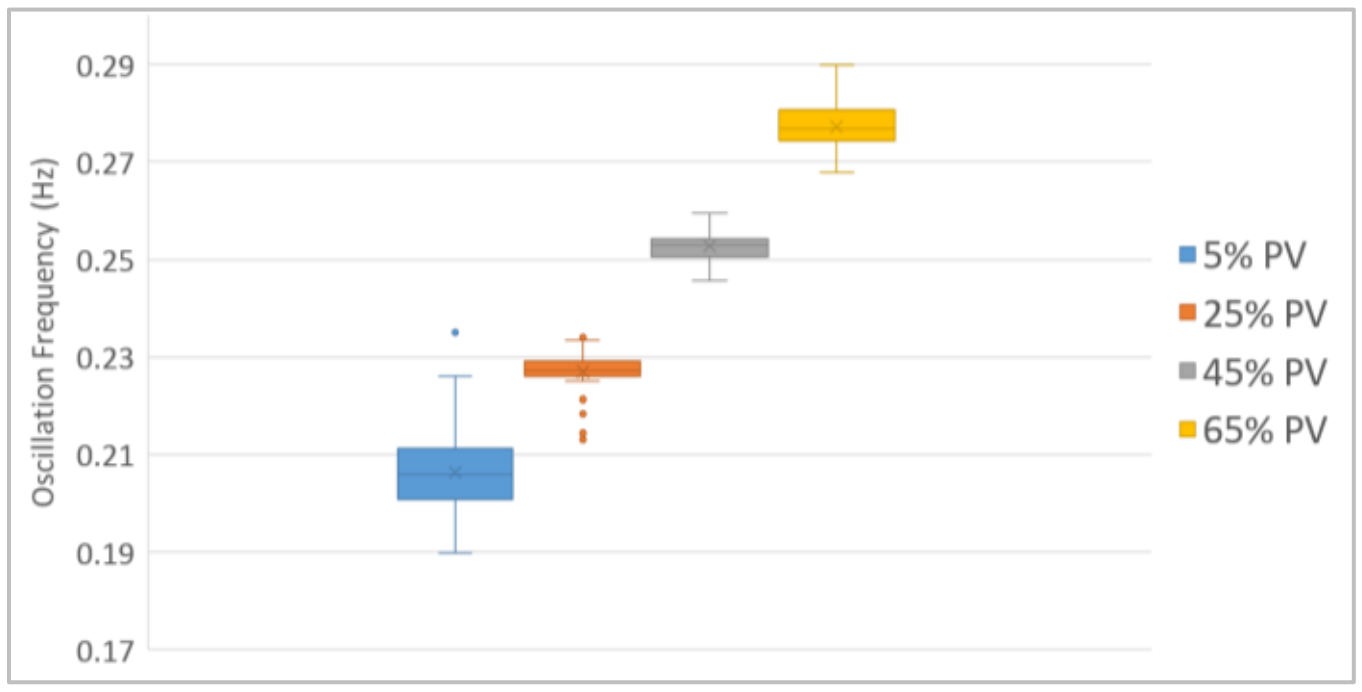

Fig. 6. Oscillation frequency change with PV penetration increases

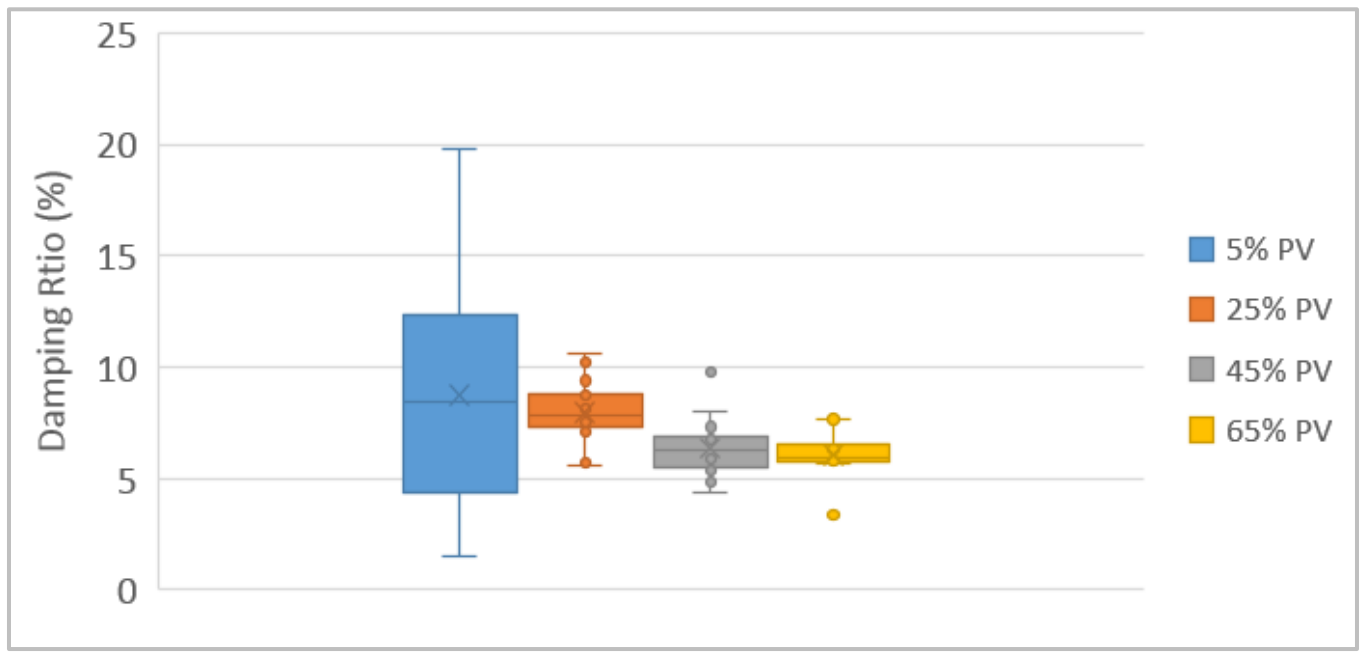

Fig. 7. Oscillation damping ratio change with PV penetration increases

As examples, Fig. 8 and Fig. 9 shows the frequency profiles of the two different locations: Connecticut (CT) and Tennessee (TN), respectively. It can be seen from Fig. 8 that the oscillation frequency increases, but the amplitude and damping decreases with PV penetration increase. In TN, while the change of this inter-area oscillation mode is similar to that of Connecticut, there is an obvious local oscillation since it is close to the disturbance location, as shown in Fig. 9. The frequency of this local oscillation is around 15 Hz for the 65% PV penetration scenario. It can be noted that the oscillation frequency increases with the PV penetration, indicating this oscillation mode change is related to the reduction of system inertia.

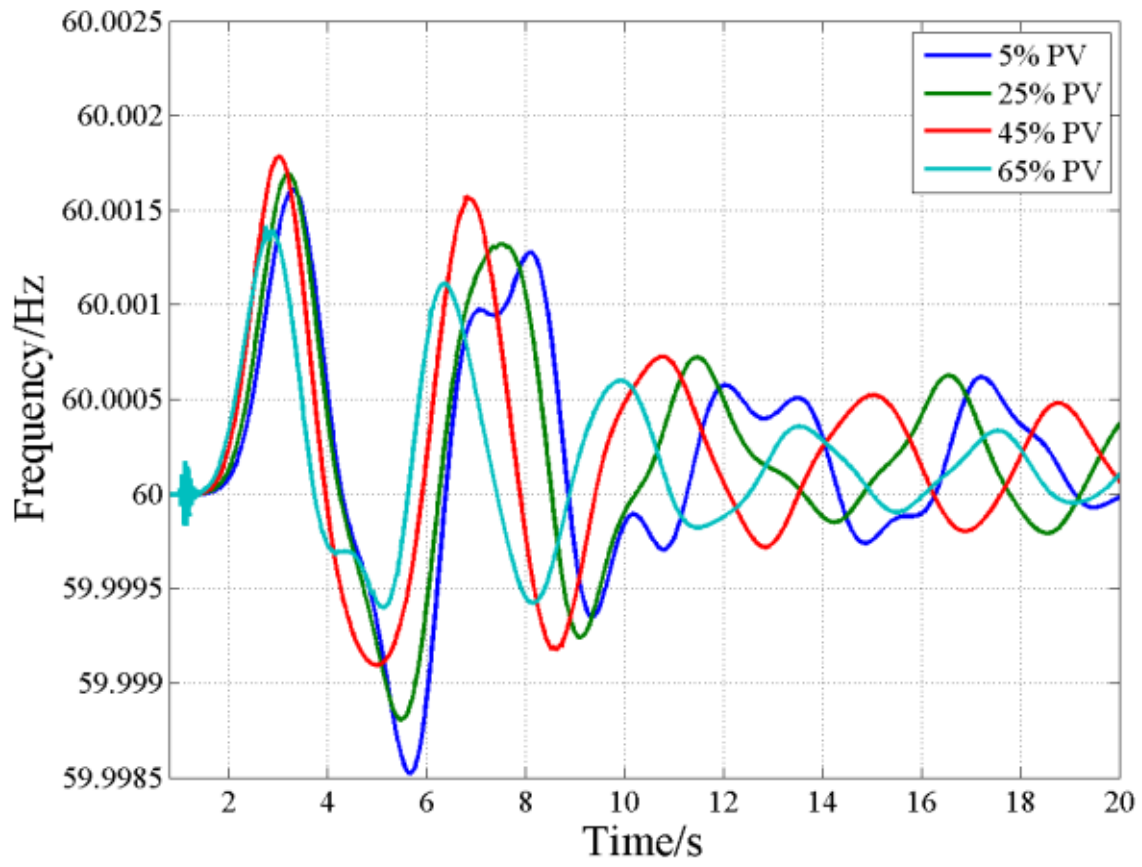

Fig. 8. Oscillation frequency change with PV penetration increases (CT)

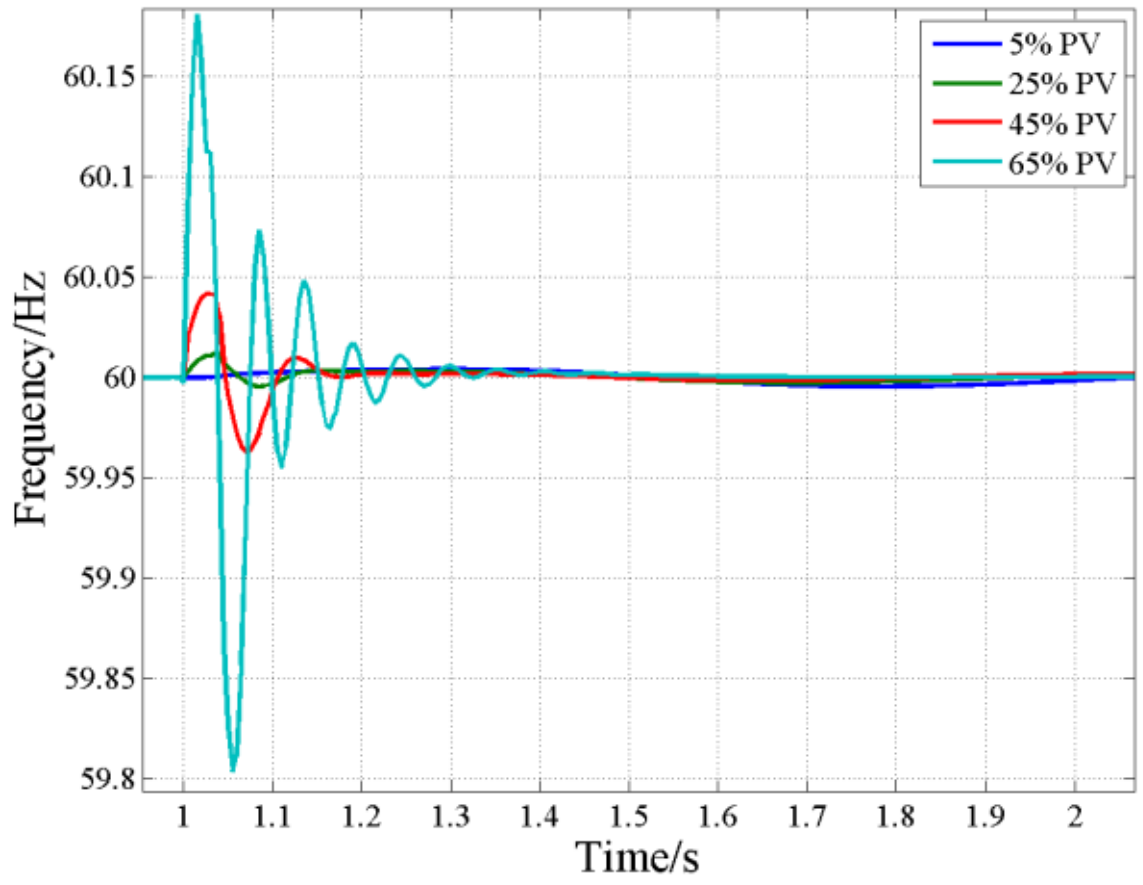

Fig. 9. Local oscillation changes with PV penetration (TN)

## B. *Impact on Mode Shape*

The mode shape describes the angular information of an oscillation mode and can be used to design additional damping controllers. Fig. 10 shows the oscillation mode under different PV penetration levels. It can be seen that the 0.2Hz inter-area oscillation mode has two main coherent groups: Northeastern EI (NY, PA, CT, and NJ, etc.) and West EI (MB, MN, and KS, etc). With the increase of PV penetration, the angular different of buses within each group decreases, indicating the increase of coherence between generators within each group due to the decrease of conventional generation.

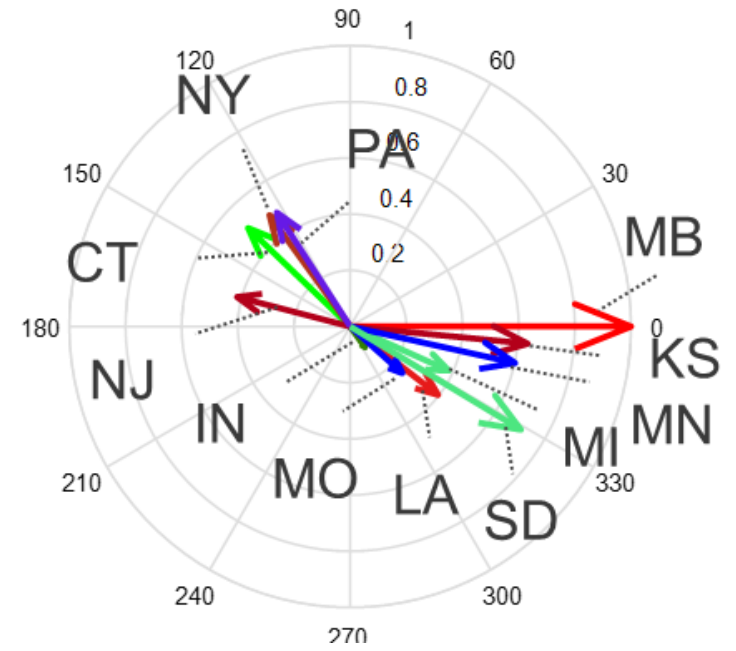

(a) 5% PV

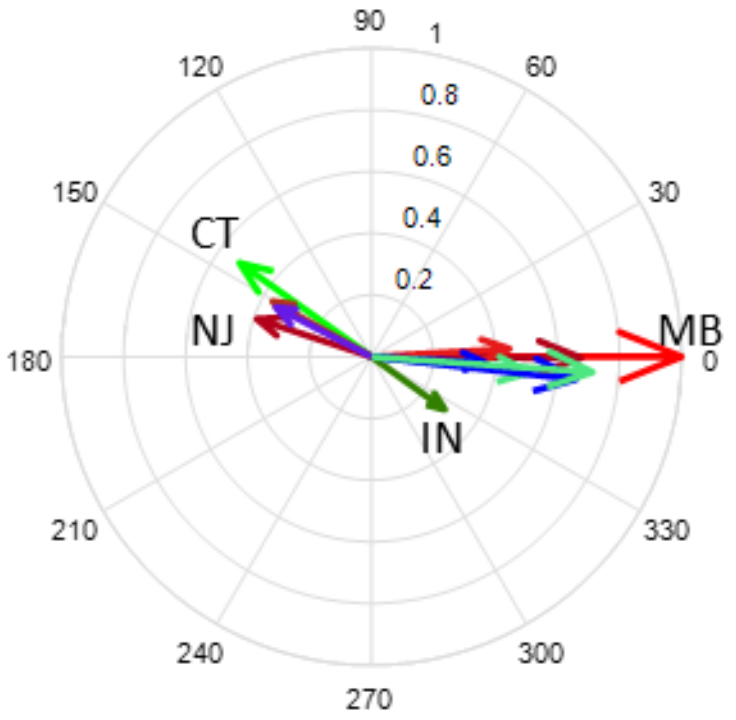

(b) 25% PV

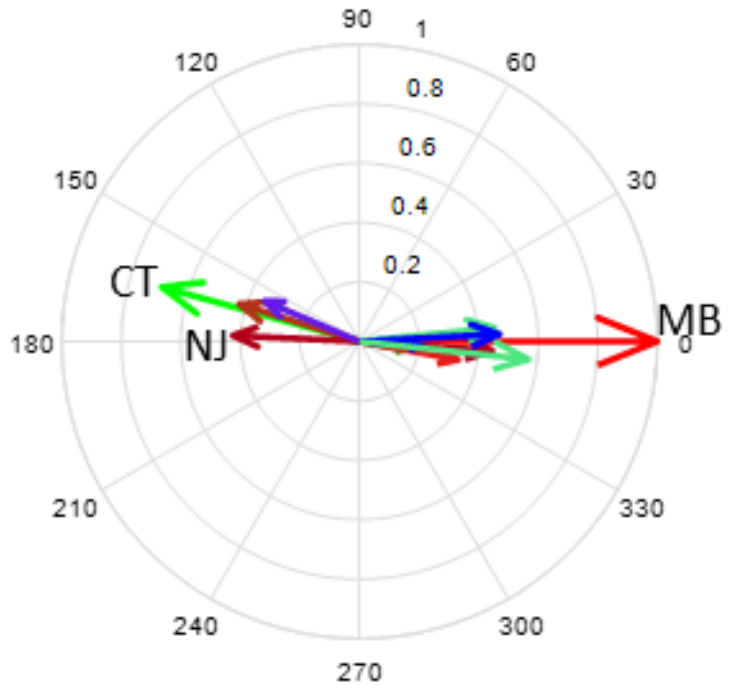

(c) 45% PV

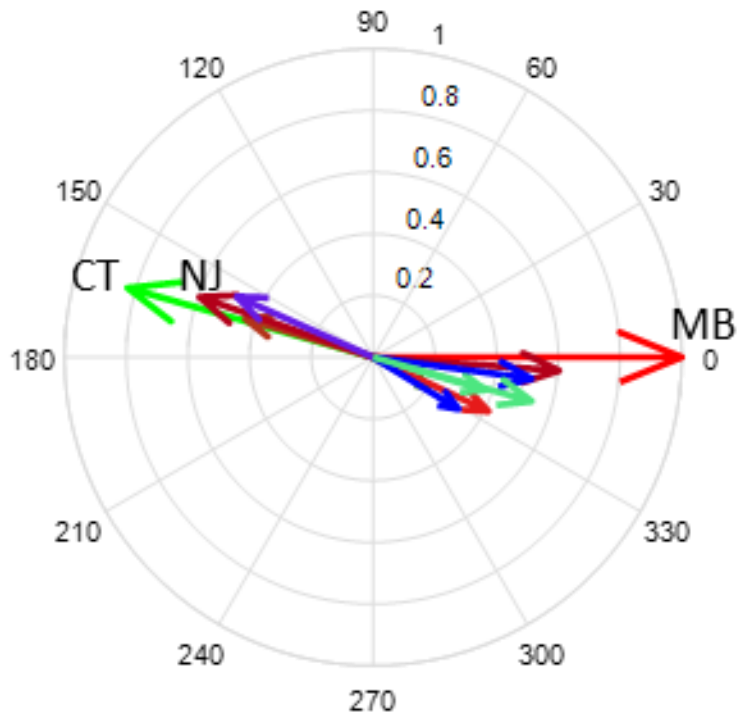

(d) 65% PV

Fig. 10. Mode shape change with PV penetration

## C. *Impact of the Control Strategy of PV Power Plants on Inter-Area Oscillation*

The control strategy of the PV power plants has impact on the inter-area oscillation mode. As a comparison of the first control strategy, Control Strategy 2 (Volt/Var control without SolarControl) is applied to all PV plants. Fig. 10 shows the frequency profile of a bus in CT under two control strategies in the 65% penetration scenario. It can be seen that the system oscillation frequencies under the two control strategies are close, but Control Strategy 2 has a smaller oscillation amplitude and damping ratio, due to the slower reset of reactive power from PV plants after disturbance.

Fig. 12 shows the oscillation mode shape under Control Strategy 2. It can be seen that the oscillation mode shape also includes two separate groups, similar to Control Strategy 1 (Fig. 10 d). but with some differences on the phase angles, especially at grid edges such as New Jersey and Manitoba, CA. This oscillation mode shape difference is due to the discrepancy in reactive power output under two PV control strategies.

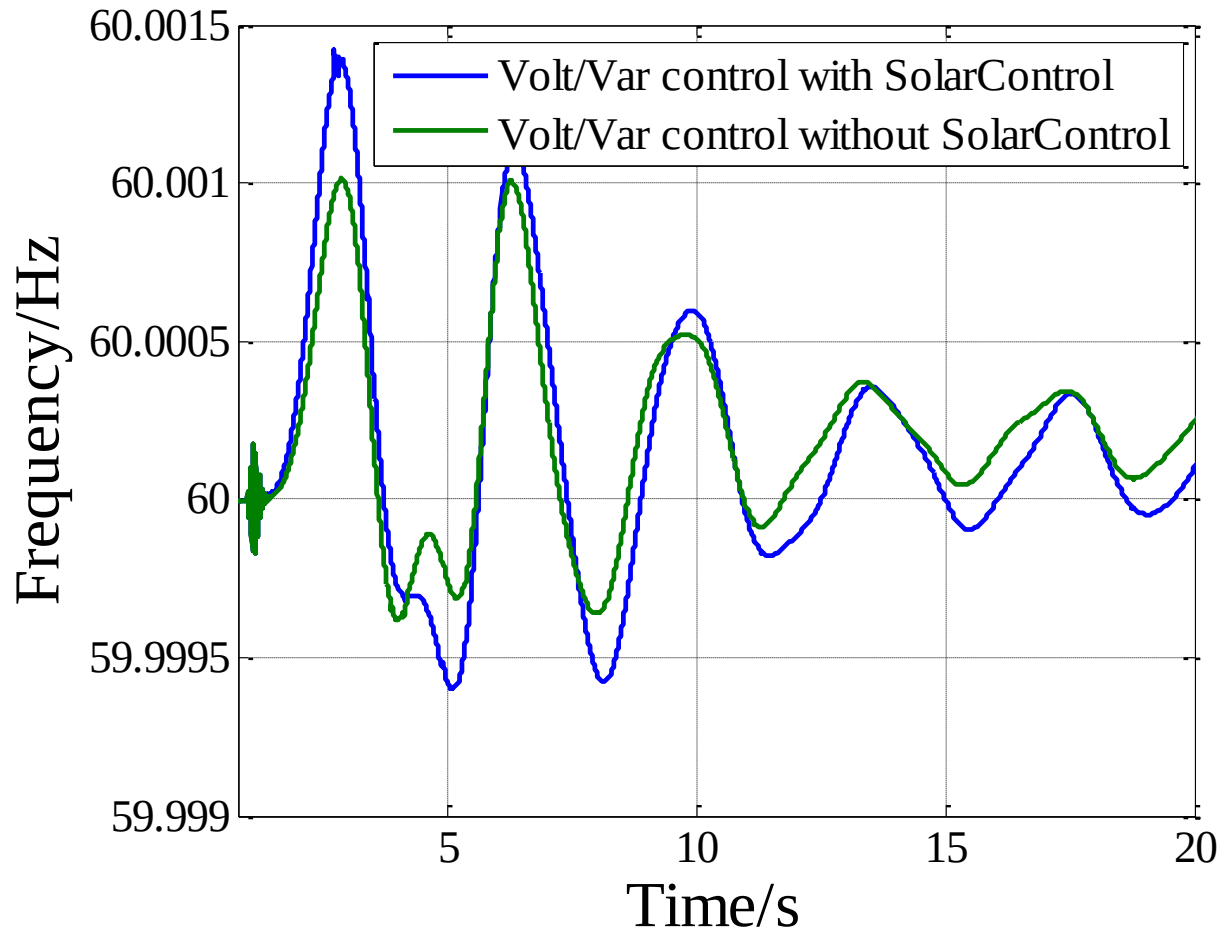

Fig. 11. Oscillation frequency in CT under different control strategy of PV power plants (65% PV)

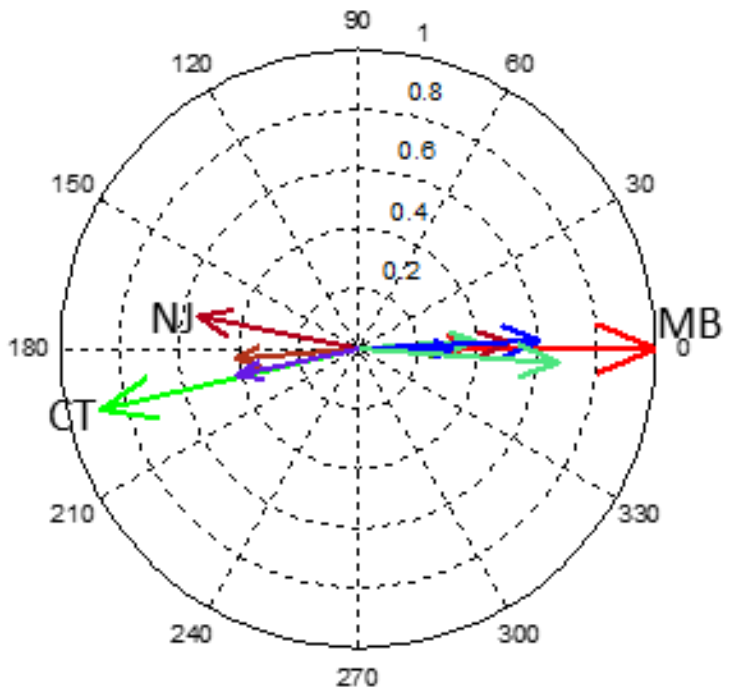

Fig. 12. Oscillation mode shape at 65% PV (Control Strategy 2)

### *D. Introduction of New Inter-Area Oscillation Modes under Certain Control Parameter Settings of PV Plants*

Certain control parameter settings can bring about new oscillation modes. As the typical setting for fast reactive power control, the reactive power regulator gain is increased to 0.5 from 0.1 (the initial value) in Control Strategy 1. Applying the same disturbance, the frequency profiles at Illinois (IL) for the 65% PV scenario under the two different settings are shown in Fig. 13. A 1.2 Hz inter-area oscillation mode can be easily seen after the disturbance for the case with a higher reactive power gain. Fig. 14 shows the mode shape of the 1.2 Hz oscillation mode. It can be noted that large oscillation locations include MI, IL, and VA, etc.

Fig. 15 and Fig. 16 shows the changes of frequency and damping ratio of this new oscillation mode for multiple PV penetration levels. Different from the 0.2 Hz mode, the frequency and damping ratio of the 1.2 Hz mode barely changes with PV penetration. This insensitivity is due to that this oscillation mode is caused by the PV controllers, which is not influenced by the time constants relevant to system inertia.

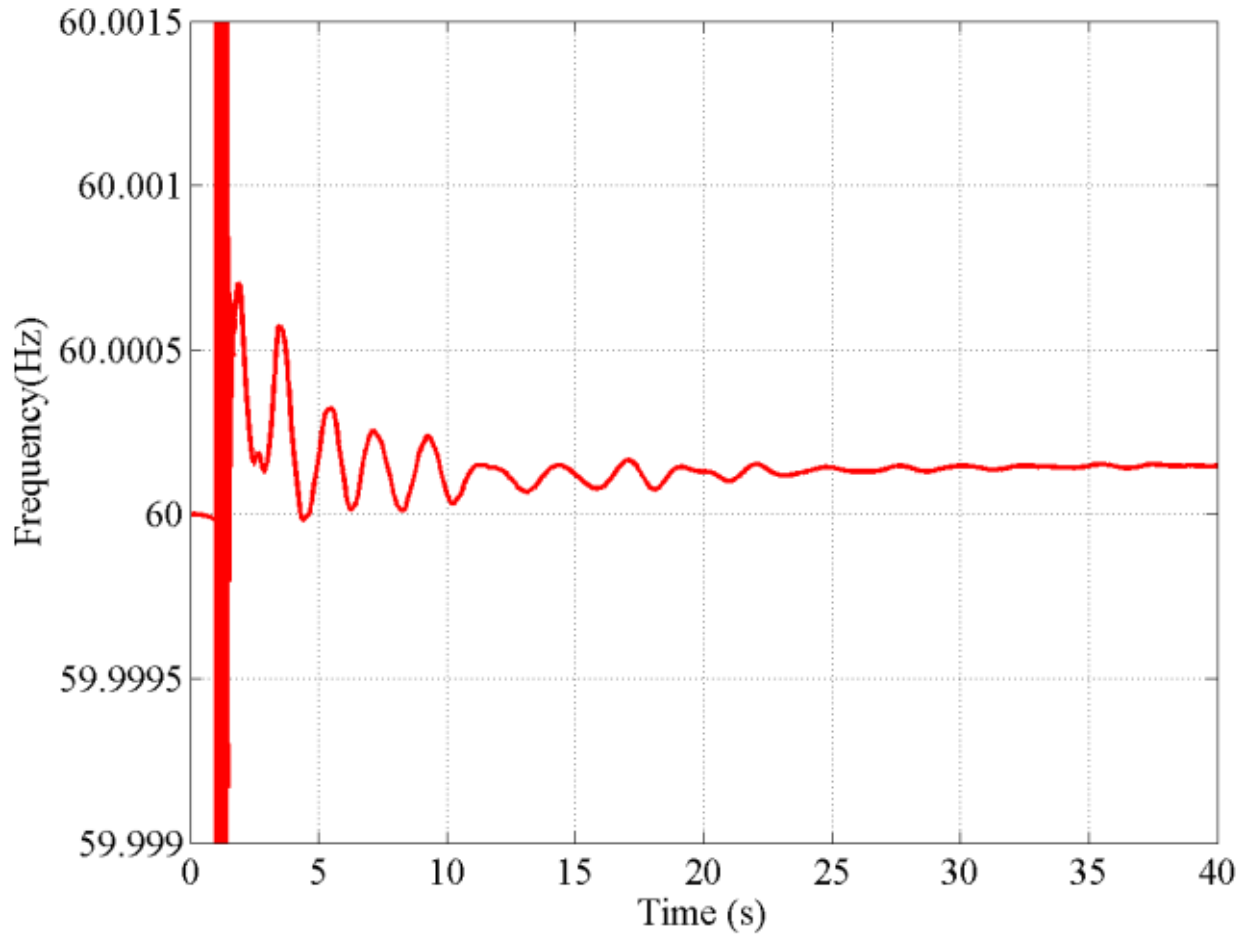

(a) Control Strategy 1 (Volt/Var control with SolarControl)

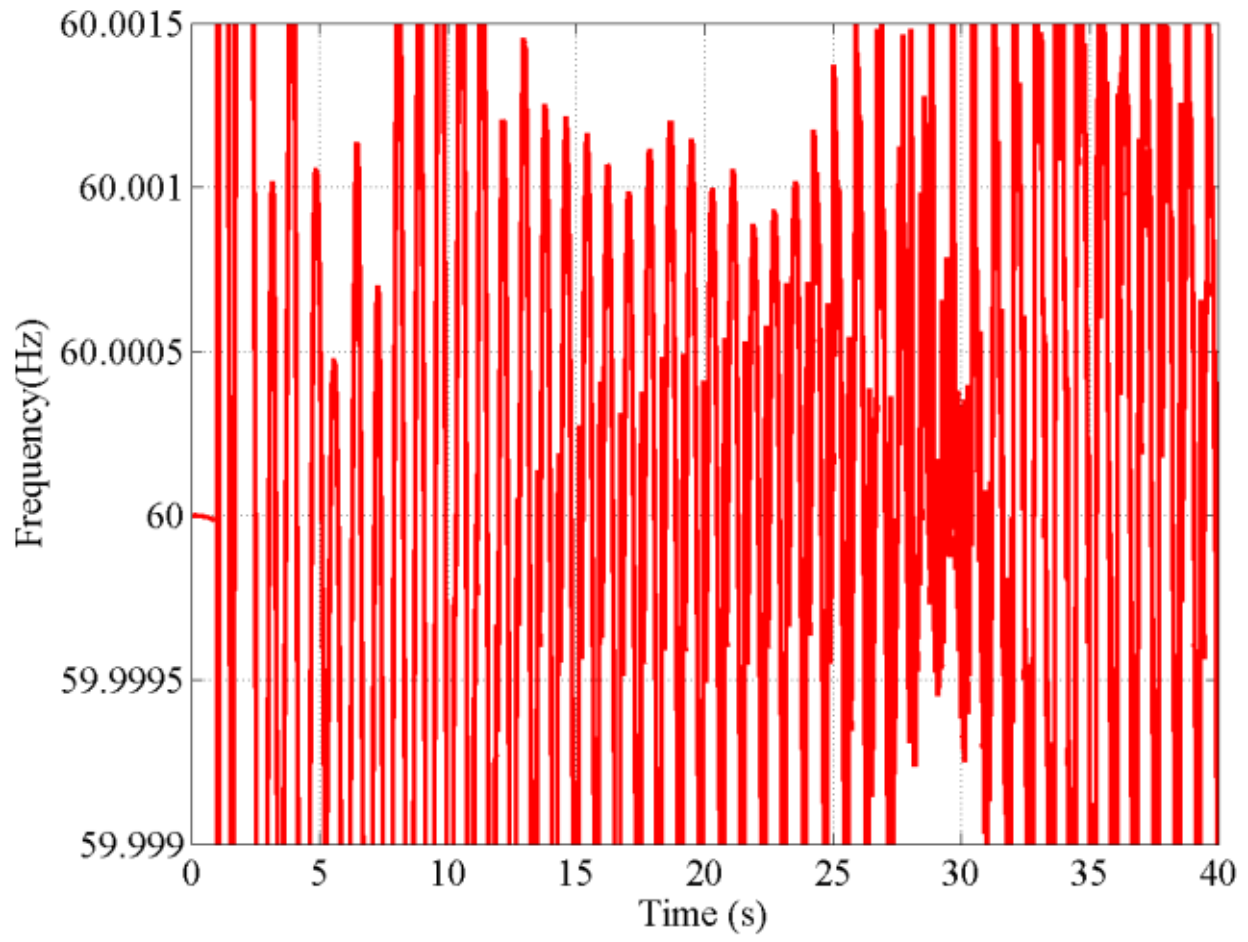

(b) Control Strategy 1 with fast power factor control

Fig. 13. Frequency profile at IL under different PV control paramter settings (65% PV)

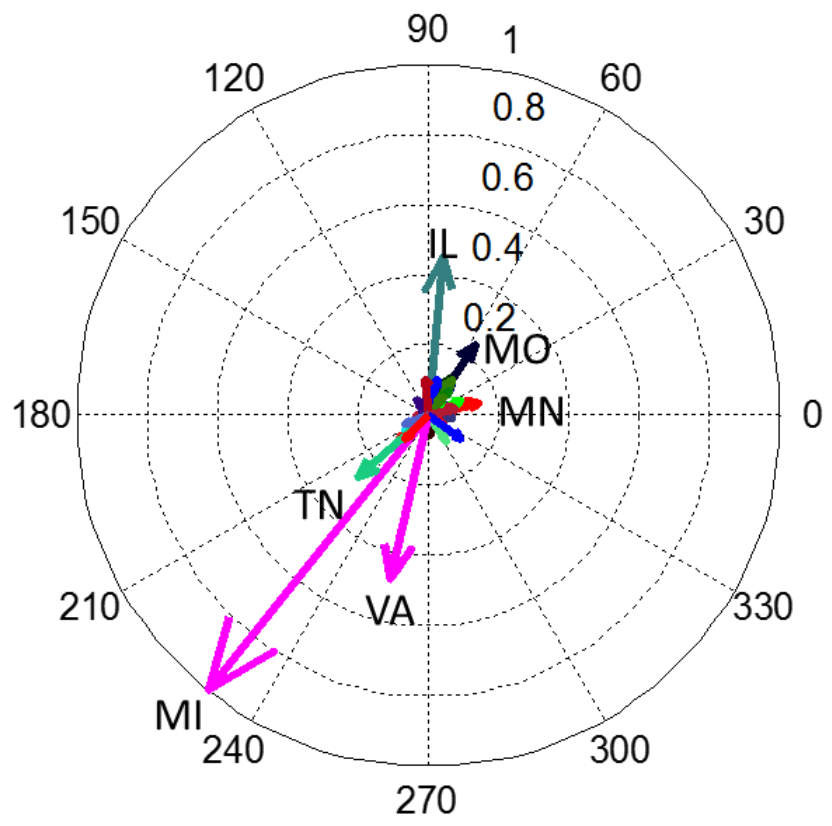

Fig. 14. Mode shape of the 1.2 Hz Inter-area mode (65% PV)

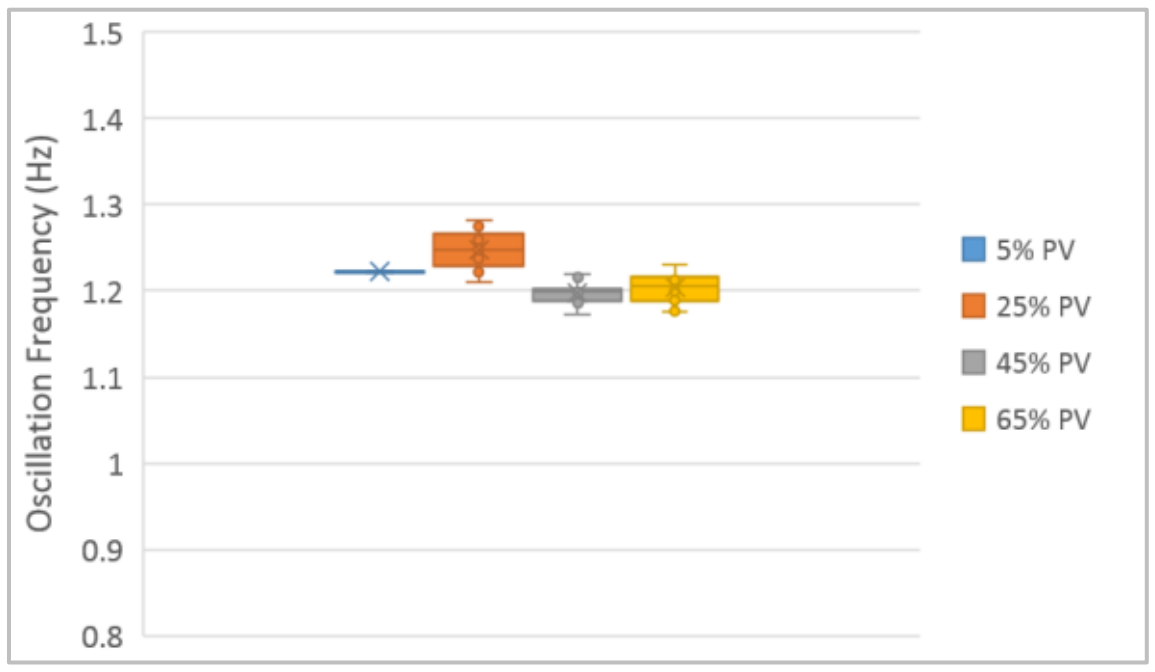

Fig. 15. Oscillation frequency change with PV penetration (1.2 Hz mode)

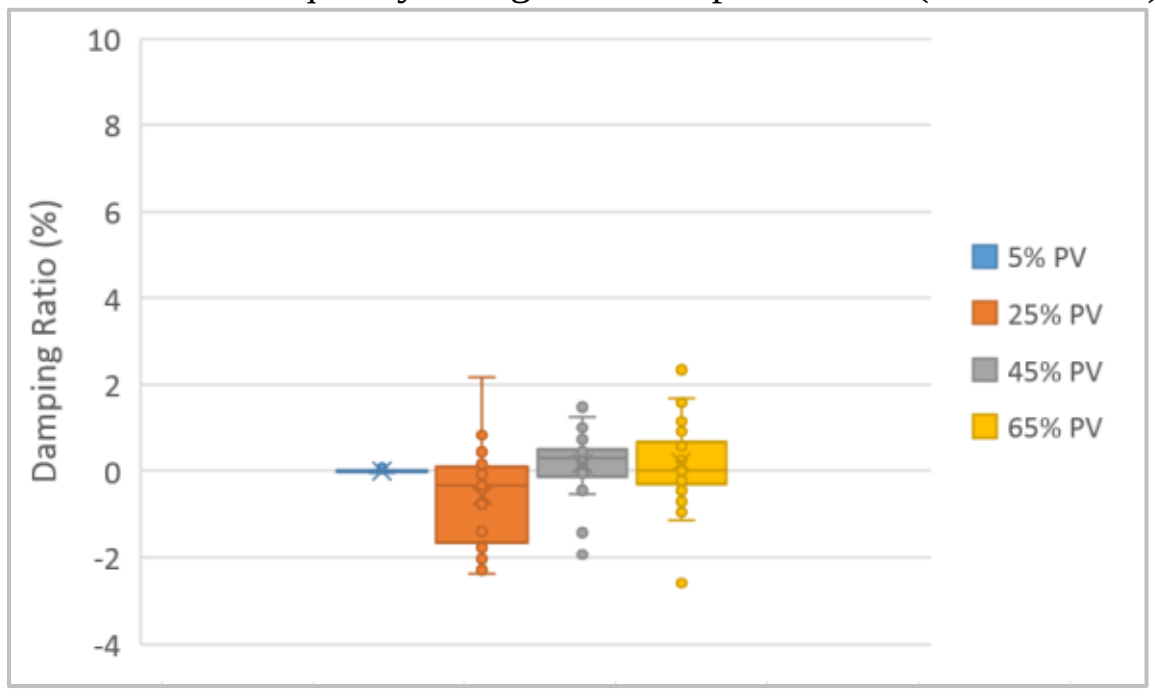

Fig. 16. Damping ratio change with PV penetration (1.2 Hz mode)

## V. Conclusions

Based on detailed system dynamic model, this paper studied the impact of high PV penetration on inter-area oscillation modes of the U.S. EI system. The analytical procedures include dynamic model development, PV plant siting, time-domain transient simulation, and Matrix Pencil analysis. Changes in the oscillation frequency, damping, and mode shape, as well as the introduction of new oscillation modes, are studied by varying PV penetration levels, control strategies and parameters. It is found that the oscillation damping decreases as PV penetration increases in the EI system, indicating that additional operation and planning consideration may be required to increase the oscillation damping as PV increases in the future. Variations in PV control strategy and parameters are found to have impact on the mode shape and could also create new oscillation modes. The results could help system operators and planners understand system oscillation behaviors design mitigation methods for future high PV system. The basic analysis procedures are generic and could be applied to other large-scale power systems.